\documentclass[twocolumn,showpacs,nofootinbib,preprintnumbers,amsmath,amssymb,floatfix,prd]{revtex4}
\usepackage{epsfig}
\usepackage{dcolumn}
\usepackage{hyperref}
\begin{document}
\title{Monte Carlo sampling variant of the DSSV14 set of helicity parton densities}
%
%
\author{Daniel de Florian}
\email{deflo@df.uba.ar} 
\affiliation{International Center for Advanced Studies (ICAS), UNSAM, 
Campus Miguelete, 25 de Mayo y Francia (1650) Buenos Aires, Argentina}
\author{Gonzalo Agust\'{\i}n Lucero}
\email{glucero@df.uba.ar} 
\affiliation{Departamento de F\'{\i}sica and IFIBA,  
Facultad de Ciencias Exactas y Naturales, Universidad de Buenos Aires, Ciudad Universitaria, Pabell\'on\ 1 (1428) Buenos Aires, Argentina}

\author{Rodolfo Sassot}
\email{sassot@df.uba.ar} 
\affiliation{Departamento de F\'{\i}sica and IFIBA,  
Facultad de Ciencias Exactas y Naturales, Universidad de Buenos Aires, Ciudad Universitaria, Pabell\'on\ 1 (1428) Buenos Aires, Argentina}
\author{Marco Stratmann}
\email{marco.stratmann@uni-tuebingen.de}
\author{Werner Vogelsang}
\email{werner.vogelsang@uni-tuebingen.de}
\affiliation{Institute for Theoretical Physics, 
University of T\"ubingen, Auf der Morgenstelle 14, 72076 T\"ubingen, Germany}

\vspace*{1cm}
\begin{abstract}
We implement a Monte Carlo sampling strategy to extract helicity parton densities and 
their uncertainties from a reference set of longitudinally polarized scattering data, chosen to 
be that used in the DSSV14 global analysis. 
Instead of adopting the simplest possible functional forms for the helicity parton distributions 
and imposing certain restrictions on their parameter space in order to constrain them,
we employ redundant, flexible parametrizations and fit them to a large number of
Monte Carlo replicas of the existing data.
The optimum fit and its uncertainty estimates are then assumed to
be given by the statistical average of the obtained ensemble of replicas
of helicity parton densities and their corresponding variance, respectively. 
We compare our results to those obtained by the traditional fitting approach and to the uncertainty 
estimates derived with the robust Lagrange multiplier method, finding good agreement. 
As a first application of our new set of replicas, we
discuss the impact of the recent STAR dijet data in further constraining the elusive 
gluon helicity density through the reweighting method.
\end{abstract}
\maketitle

\section{Introduction and Motivation}
%
The precise determination of parton distribution functions (PDFs) is a key ingredient
to establish the validity and accuracy of perturbative QCD factorization and the assumed
parton density universality, and, therefore, of our current understanding of 
the nucleon structure and the strong interactions at the most fundamental level as expressed
in term of quarks, antiquarks, and gluons \cite{Collins:1989gx}. 
This is especially the case for helicity PDFs that reflect the spin alignment of quarks 
and gluons relative to that of their parent nucleon spin, ever since the milestone 
deep-inelastic scattering (DIS) experiment carried out by EMC at CERN thirty years ago \cite{ref:emc-a1p}.
The outcome challenged the naive quark model, showing that little of the proton spin 
is carried by the quarks and antiquarks. 

The EMC result was later on confirmed by similar experiments at SLAC, DESY, CERN, and JLAB, 
and complemented with semi-inclusive DIS (SIDIS) measurements in order to pin down how the 
different quark and antiquark species are polarized individually \cite{Aidala:2012mv}. In addition,
a vigorous polarized proton-proton collision program was carried out at the BNL Relativistic 
Heavy Ion Collider (RHIC) \cite{Aschenauer:2015eha} and 
established, among other things, that gluons in a polarized proton themselves carry polarization \cite{deFlorian:2014yva}.
This result gives rise to new compelling questions 
such as how much room, if any, is left for the quark and gluon orbital angular momenta in the proton spin balance, 
and what the actual correlation between spin and parton flavor is. 

In order to address these questions quantitatively, 
increasingly refined phenomenological tools to analyze, combine, and compare 
the increasing number of precise experimental results probing the nucleon spin structure
in a single, consistent, and accurate theoretical framework are required.
This is precisely the purpose of global QCD analyses, that, in the case of helicity parton distributions, 
have matured significantly in the past five years \cite{deFlorian:2014yva,Nocera:2014gqa,Sato:2016tuz}.
They routinely include complementary DIS, SIDIS, and $pp$ observables as provided by current 
experimental programs, and are prepared for the challenges of a future generation of 
measurements, namely those foreseen at an Electron Ion Collider (EIC) \cite{Aschenauer:2013iia,Aschenauer:2015ata}.
The possible realization of an EIC within the next decade is currently 
under active discussion in the U.S. \cite{Accardi:2012qut}.

The estimate of uncertainties for PDFs as obtained through a global analysis of diverse sets 
of data with different characteristics of uncertainties 
is in general \cite{Butterworth:2015oua} -- and particularly for helicity PDFs --
a formidable task. It is still a central, open problem that has been approached 
with complementary strategies with an increasing level of sophistication in the past ten years 
\cite{Stump:2001gu,Pumplin:2001ct,Martin:2009iq}. 
PDFs inherit uncertainties not only from those associated with the data, 
which are in most cases well accounted for, but also from the 
theoretical framework used to link the PDFs with the measured observables. 
The latter is necessarily based on certain assumptions, 
like adopting collinear factorization, restricting oneself to a 
leading-twist approximation, and truncating any perturbative calculation at 
a certain order in the strong coupling expansion. The errors associated with these procedures are 
extremely difficult to quantify. 
In addition, PDFs may be biased by the way in which the analysis interpolates between the values of parton 
momentum fraction probed by the data.
Traditionally this has been done by assuming a given functional form for each parton flavor
at some initial reference energy scale of order $1\,\mathrm{GeV}$. 
More recently, more elaborate procedures based on neural 
networks \cite{DelDebbio:2007ee,Ball:2008by,Nocera:2014gqa} have been invoked for this task.
Finally, global analyses usually have to make certain 
simplifying assumptions, such as flavor-symmetry relations between the (anti-)quark distributions, 
sum rules for moments,
using some model estimates for potential nuclear effects for data taken with nuclear targets,
and adopting values for heavy quark masses and other fundamental constants and parameters. 
It is highly nontrivial how these approximations and choices eventually propagate into the obtained 
set of PDFs and into estimates for cross sections and other experimental observables computed with them.

The assessment of uncertainties for helicity PDFs has evolved from a mere
comparison between sets based on very simple parameterizations for their functional form 
and different -- albeit at that time equally plausible -- simplifying assumptions on
the available parameter space, 
to a rather sophisticated combination of Monte Carlo samplings of the data combined with neural 
network techniques as a largely unbiased interpolating strategy \cite{Ball:2013lla,Nocera:2014gqa}. 
Also, the traditional approach to determine helicity PDFs based on more 
restrictive parameterizations, which allows for numerically very efficient computations 
of arbitrary observables in Mellin space \cite{Stratmann:2001pb} at any desired order in perturbation theory, 
has been supplemented with the improved Hessian technique \cite{Pumplin:2001ct} and the robust
Lagrange multiplier method \cite{Stump:2001gu} to estimate and propagate uncertainties more reliably
\cite{deFlorian:2009vb}. Alternatively, other ideas have been pursued by implementing 
iterative Monte Carlo fitting techniques, that combine Monte Carlo sampling of the 
parameter space with a resampling of data and cross validation methods \cite{Sato:2016tuz}.

A common and often controversial feature of both the improved Hessian approach and
the use of the Lagrange multiplier method to estimate uncertainties is the necessity 
to introduce a suitable tolerance criterion, usually obtained by inspection of the quality of
the fit to all the available sets of data. 
This has to be done to accomplish sensible error estimates that, 
for example, fulfill the naive expectation that the majority of the data in the fit fall 
within the quoted uncertainty bands. 
These tolerances typically imply increments $\Delta \chi^2$ of the 
effective $\chi^2$ function used to measure the quality of the fit between theory and experiment 
substantially larger than the textbook expectation, $\Delta \chi^2=1$, for the 
68\% confidence level (C.L.) interval. 
It is quite apparent that any global PDF fit, apart from the bias from choosing a functional form
and the often neglected correlations between the parameters of the fit and/or within the data, 
is very far from the idealized case that leads to the criterion of $\Delta \chi^2=1$. 
It also suffers from several sources of errors inherent to the theory approximations 
that cannot be properly accounted for, are highly non-Gaussian, and, in any case, are usually neglected. 
Fitting only a single type of observable, say, just DIS data, PDFs may effectively compensate for 
or hide such defects, but this is much less likely in any truly global fit to measurements 
with rather diverse characteristics such as DIS, SIDIS, and various sets of $pp$ data.

Monte Carlo sampling strategies to obtain PDFs \cite{DelDebbio:2007ee,Ball:2008by}, on the other hand, 
avoid the adoption of a tolerance criterion and some other shortcomings in
the propagation of PDF uncertainties to experimental observables. 
Schematically, one first generates a Monte Carlo ensemble 
of replicas of the original data with a probability distribution derived from the 
reported errors within the desired accuracy. In a second step, a PDF set is obtained 
for each replica of the data. The so obtained ensemble of corresponding PDF replicas 
is expected to contain all the information relevant for the PDF determination: 
the central value of a PDF, or any quantity derived from them, is taken to be the average 
over the PDFs replicas, and the corresponding uncertainty is the statistical standard deviation. 
Interestingly, the uncertainty estimates for unpolarized PDFs
derived in this way are fairly consistent with those obtained with the traditional 
approach when similar data sets are fitted and theoretical approximations are made,
provided a substantially larger tolerance $\Delta \chi^2\gg 1$ is adopted for the latter
\cite{Butterworth:2015oua}.
Such an agreement would suggest that the ensemble 
of PDF replicas can account for the arguments used to motivate such large 
tolerance criteria, but effectively avoids the arbitrariness of defining 
a specific $\Delta \chi^2$ and has a much clearer statistical interpretation.  
  
In the following, we will implement a Monte Carlo sampling strategy to determine helicity 
PDFs. To this end, we study the same set of polarized scattering data utilized in the well-known and frequently used 
DSSV14 analysis \cite{deFlorian:2014yva}, a global fit at next-to-leading order (NLO) accuracy 
of DIS and SIDIS data together with results on the hadroproduction of jets and neutral pions in polarized proton-proton
collisions from BNL-RHIC. The DSSV14 analysis was based on the more traditional fitting methodology, 
and uncertainties were estimated with the Lagrange multiplier technique \cite{deFlorian:2014yva}
and an appropriately chosen tolerance $\Delta \chi^2\gg 1$.
In addition, we also adopt identical theory inputs and conventions to facilitate the comparison 
between both approaches and their results. As in the case of unpolarized PDFs, and as we shall demonstrate below, 
we find a rather good agreement between the central values and variances coming from
the newly derived Monte Carlo replicas and the best fit and 68\% C.L.\ uncertainties from the DSSV14 analysis.
   
Along with a detailed comparison between the two methods, 
we also provide a large set of PDF replicas, available upon request from the authors,
which are representative of the uncertainties of the original DSSV14 analysis, but 
much easier to apply to any desired observable than the Lagrange multiplier method. 
Hence, the replicas may be straightforwardly 
employed to estimate PDF uncertainties in any new or forthcoming future measurement, 
as well as to include information from data sets not yet included in the original DSSV14 fit 
by means of a reweighting technique \cite{Ball:2010gb,Ball:2011gg}. 
As a first example, we shall analyze in this way the impact of recent dijet data obtained by the STAR
experiment \cite{Adamczyk:2016okk,Adam:2018pns} on the determination of helicity PDFs at NLO accuracy.
Finally, the set of replicas will be also particularly useful for a comparison with the forthcoming update 
of the DSSV14 analysis, comprising all the data that have become available since the original fit, 
as well as new theoretical inputs such as updated unpolarized PDFs and fragmentation functions.

The remainder of the paper is organized as follows.
In the next section, we very briefly remind the reader of the main aspects of the DSSV14 analysis,
the data sets included in the fit, the parameterizations assumed for the helicity
PDFs, and other relevant theoretical inputs. We also describe the implementation of the Monte Carlo
sampling of the data to obtain our set of helicity PDF replicas based on much more flexible functional forms. 
Next, in Sec.~III, we discuss the main properties of our large set of replicas and compare the results for
the individual helicity PDFs and their uncertainties to those from the DSSV14 analysis based 
on the Lagrange multiplier method. In Sec.~IV, we present a reweighting exercise 
based on recent data on dijet production in polarized proton-proton collision obtained by the STAR collaboration as 
a first example of the usefulness of the new approach. We summarize the main findings in Sec.~V. 

\section{DSSV Analysis, Parameter Space, and Monte-Carlo Sampling of Replicas \label{sec:setup}}
The DSSV14 analysis \cite{deFlorian:2014yva} is a traditional NLO extraction of helicity parton 
distributions obtained from inclusive and semi-inclusive lepton-proton, lepton-deuteron 
and lepton-helium DIS data \cite{Aidala:2012mv}, together with single-inclusive, 
high transverse momentum jet and neutral pion production measurements in polarized proton-proton collisions at RHIC \cite{Aschenauer:2015eha}. 
DSSV14 is actually an upgrade to DSSV08 \cite{deFlorian:2008mr,deFlorian:2009vb}, 
the first truly global NLO QCD analysis of spin-dependent
PDFs that combined DIS and SIDIS data with early results from RHIC,
conceived to incorporate crucial, new experimental information that appeared after 2008.
Most importantly, the DSSV14 analysis revealed for the first time evidence of a nonvanishing 
polarization of the gluons in the proton \cite{deFlorian:2014yva} at medium momentum fractions, 
a result that was later on confirmed in an independent analysis \cite{Nocera:2014gqa} based on 
the neural network approach and the reweighting technique. 

The main features of the DSSV analyses, the selection of data sets,  
details on the computation of the observables using the numerically efficient 
Mellin transform techniques, the minimization strategy, and the 
uncertainty estimates utilizing both the improved Hessian approach and the Lagrange multiplier technique
have been discussed at length in the literature and can be found in \cite{deFlorian:2008mr,deFlorian:2009vb,
deFlorian:2014yva} and references therein. 
Here, we just briefly remind the reader of the aspects that are most essential for the present analysis
and that will be altered as we proceed.

All DSSV analyses so far have adopted the traditional approach at NLO accuracy
outlined in the Introduction and set out by assuming a
flexible functional form to parameterize the helicity PDFs as functions of the parton momentum fraction $x$ 
at an initial scale of $\mu_0=1\,\mathrm{GeV}$,   
\begin{equation}
\label{eq:pdf-input}
x\Delta f_i(x,\mu_0) = N_i\, x^{\alpha_i} (1-x)^{\beta_i}
(1+\gamma_i \sqrt{x}+\eta_i x^{\kappa_i})\,,
\end{equation}
where the label $i$ denotes different flavor combinations $\Delta u+\Delta \bar{u}$, 
$\Delta d+\Delta \bar{d}$, $\Delta \bar{u}$, $\Delta \bar{d}$, $\Delta \bar{s}\equiv\Delta s$, 
and the gluon density $\Delta g$. As usual, $\Delta f_i$ represents the difference of densities
with parton spins aligned and anti-aligned with the longitudinal parent proton's spin.
The optimization of the fit
to data is carried out by varying the set of fit parameters $\{a_i\}=\{N_i,\alpha_i,\beta_i,\gamma_i,\eta_i,
\kappa_i\}$ iteratively as long as a minimum in the effective $\chi^2$ function is reached.
In each iteration the PDFs are evolved to the scales $\mu>\mu_0$ relevant in experiment 
and used to compute the corresponding observables and the effective 
$\chi^2$ function to be minimized. 

In practice, however, the currently available data do not fully constrain the entire $x$-dependence 
allowed by the parameterizations in Eq.~(\ref{eq:pdf-input}). A whole range of values 
for the fit parameters $\{a_i\}$ leads to equally good fits. 
Therefore, in the standard minimization approach, some restrictions on the parameter 
space have to be imposed such that a unique and stable minimum in $\chi^2$ can be found, 
provided that the obtained optimum value for $\chi^2$ per degree of freedom, $\chi^2/\mathrm{d.o.f}$,
does not deteriorate significantly. 
For instance, in the DSSV analyses no improvement in the quality of the best fit is found 
by allowing the parameter $\gamma_i$ to be different from zero for both the sea quarks and the gluon.
Also, $\kappa_i$ different from unity only has some impact for the gluon density $\Delta g$ but
not for any of the quark flavors.
Along the same lines, the parameters $\beta_{i=\bar{u},\bar{d},\bar{s}}$, that 
determine the large-$x$ behavior of sea quark helicity distributions, are only very weakly constrained 
by the existing data and are mostly affected by the positivity condition, 
$|\Delta f_i| \le f_i$, relative to a chosen set of unpolarized PDFs $f_i$. 
Hence, they are set to a common, fixed value within the 
positivity constraint. 
Finally, the small-$x$ behaviors of $\Delta \bar{u}$ and $\Delta \bar{d}$,
controlled by $\alpha_{i=\bar{u},\bar{d}}$ in Eq.~(\ref{eq:pdf-input}), can be tied 
to those of $\Delta \bar{u}$ and $\Delta \bar{d}$, respectively, with no detrimental effects on
the obtained $\chi^2/\mathrm{d.o.f}$. Likewise, no improvement of the fit is 
found by allowing $\alpha_{\bar{s}}$ to be different from $\alpha_{\bar{d}}$.

Although the restrictions on the parameter space listed in the preceding paragraph do not
undermine the quality of the best fit as measured in terms of $\chi^2/\mathrm{d.o.f}$, 
they certainly restrict the possible range of variation of the distributions away from the best fit 
in some uncontrolled fashion. To explore uncertainties of helicity PDFs reliably, 
any such restriction on the parameter space has to be released, for instance, 
in the implementation of the Lagrange multiplier method or when obtaining
an ensemble of replicas of the PDFs. This is precisely what we do in the following.
Since the replicas will be determined by fitting corresponding replications of the data
that individually could show features different from those that were found in the DSSV analyses 
of the actual data sets, the extra freedom of an unrestricted 
parameter space in Eq.~(\ref{eq:pdf-input}) could also matter.

Of particular interest in the quest of understanding the proton spin structure quantitatively 
is the gluon helicity density $\Delta g$, which, however, turns out to be
the least known distribution. It is constrained mainly by RHIC data, in a restricted
range of momentum fractions $x$ \cite{deFlorian:2014yva} though and, to a much lesser extent,
by relatively suppressed NLO corrections to the DIS and SIDIS cross sections 
and, indirectly, through the scale dependence of the parton distributions.
In face of that, in our present analysis we allow for an even more redundant functional form for $\Delta g$
than Eq.~(\ref{eq:pdf-input}), in order to maximize the decoupling between the so far 
weakly constrained low and high-$x$ regions: 
\begin{eqnarray}
\label{eq:glue-input}
x\Delta g(x,\mu_0) =\!\!\!\!\!\!&&N_g x^{\alpha_g} (1-x)^{\beta_g}\nonumber\\
&&\times
\left(1+\eta_g x^{\kappa_g}\right) \left[1+ \delta_g\, x^{\rho_g} (1-x)^{\theta_g} \right]\,.
\end{eqnarray}

Such a proliferation of fit parameters $\{a_i\}$ in Eqs.~(\ref{eq:pdf-input}) and (\ref{eq:glue-input})
requires a very exhaustive sampling strategy of the parameter space to obtain the replicas reliably. 
Our Monte Carlo sample of replicas of the experimental data is generated as a multi-Gaussian
distribution. For each data point for a measured spin asymmetry
$A^{\mathrm{(exp)}}_i$ corresponding to a specific observable and 
kinematics, we generate 1000 replicas, labeled by a superscript $(k)$, as follows \cite{Ball:2010gb}
\begin{eqnarray}
\label{eq:replica-input}
A^{\mathrm{(rep)}(k)}_i \!\!&=&\!\! \left(1+ r_N^{(k)}\sigma_{N} \right) \nonumber \\
&&\times \left( A^{\mathrm{(exp)}}_i + \sum_{p=1}^{N_{\mathrm{sys}}} r_p^{(k)} 
\sigma_{i,p}+ r_i^{(k)}\sigma_{i,s} \right), 
\end{eqnarray}
with $r^{(k)}$ denoting independent, univariate Gaussian random numbers 
for each independent source of errors \cite{DelDebbio:2007ee}. 
$\sigma_N$ stands for the global normalization error of a data set, 
while $\sigma_{i,p}$ and $\sigma_{i,p}$ are of systematical and statistical origin, respectively. 
These errors include the statistical and systematic errors of the measurements
reported by the experiments, and could in principle also include estimates of those
stemming from the theoretical inputs used to compute the observables, such as the
fragmentation functions needed for SIDIS spin asymmetries and the choice
for the unpolarized PDFs appearing in the denominators of the spin asymmetries.
In practice, the latter are usually considered subdominant and neglected
\cite{Nocera:2017wep}. However, we include an estimate of the uncertainty 
related to the FFs as a further (theoretical) error when computing the contribution of the
SIDIS data to $\chi^2$ and when generating the replicas. Technically, we do this 
with the help of the Hessian uncertainty sets of Refs.~\cite{deFlorian:2014xna,deFlorian:2017lwf}.
Any other theoretical errors associated with the truncation of the perturbative expansion at NLO accuracy, 
the value of the strong coupling, potential nuclear corrections or higher twist contributions, 
and possible departures from the collinear approximation are ignored.

One should recall that the data included in the fit \cite{deFlorian:2014yva}, and, 
consequently, their replicas, span a much
more limited range of parton momentum fractions $x$ than in the case of unpolarized PDFs. In fact,
only a handful of DIS data points reach below $x\sim 10^{-2}$, albeit in a 
very restricted range of photon virtualities ($1.1 \le Q^2 \le 2.1$ GeV$^2$) which sets
the relevant energy scale of the DIS process. Similarly, RHIC $pp$ data are only sensitive
to momentum fractions $x\gtrsim 5\times10^{-3}$ but at somewhat higher energy scales
set by the transverse momentum of the observed pion or jet.
This leads to an extremely poor constraint on the helicity PDFs in general, and $\Delta g$ in
particular, in the range $10^{-3} \lesssim x\lesssim 10^{-2}$, and leaves 
them completely undetermined below, despite some indirect constraints from the
positivity requirement, i.e., through the steep rise of unpolarized PDFs at small-$x$,
and the integrability condition. The latter states that the first moments of helicity PDFs, 
$\int_0^1 \Delta f_i(x,Q^2)\,dx$, must not diverge, as they express the net contribution 
of a given parton flavor to the spin of the proton. 

At variance with an implementation of the Monte Carlo sampling of data combined with a
neutral network description of the momentum dependence of PDFs \cite{Ball:2013lla,Nocera:2014gqa}, 
that avoids a potential cross-talk between different regions of $x$, 
data in the measured $x$-range can induce a fictitious behavior in the
unmeasured region when the fitting is performed based on a given functional form. 
For instance, the improvement in the quality of the fit to data at higher values of $x$  
can be infinitesimally small but it might still drive the behavior of PDFs 
in the unmeasured $x$-regime and invalidate uncertainty estimates 
due to some residual rigidity in the low-$x$ extrapolation.
Therefore, any results in low-$x$ region, i.e.\ below about $x\simeq 10^{-3}$,
should not be considered as faithful or stemming from any existing experimental result. 

Since many current feasibility and impact studies for a future EIC \cite{Accardi:2012qut},
see, for instance, Refs.~\cite{Aschenauer:2015ata,Ball:2013tyh}
are highly interested in exploring the uncharted low-$x$ domain of helicity PDFs, 
we will provide extrapolations of our ensemble of replicas 
beyond the kinematic region where data faithfully constrain them.
To this end, we supplement the Monte Carlo
sampling approach with information coming from the Lagrange multiplier method.
The latter allows one to estimate the uncertainty of any observable dependent on the PDFs 
or of the PDFs themselves within any given confidence level limit 
and under the assumption of a given functional form.
We use this extra information to generate a set of 10 pseudo-data points 
uniformly distributed in logarithmic scale between $x=10^{-6}$ and $x=10^{-3}$,
i.e., outside the range spanned by actual data,
with a Gaussian error distribution around the result of the DSSV14 
best fit for the gluon helicity distribution,  with variances corresponding to the 68\% 
CL limit estimated in the Lagrange multiplier method as discussed below.
We have checked, of course, that this addition does not modify the results 
in the region of validity of the Monte Carlo sampling method.
One should always keep in mind that the so obtained low-$x$ extrapolations of the helicity PDF replicas 
have some explicit parameterization bias and are 
solely provided for the purpose of feasibility and impact studies for a future EIC. 

The gist of the Lagrange multiplier method \cite{Stump:2001gu} is to
study the behavior of the effective $\chi^{2}$ measure used to quantify the goodness of 
a global fit as a function of the fit parameters $\{a_i\}$ or, alternatively, for
any observable $\textit{O}(\{a_i\})$ of interest depending on them. Most importantly,
there is no need to assume anything about the $\chi^{2}$ function in the vicinity of its
minimum or any relations between the fit parameters and observables. 
Schematically, the method is implemented in practice by minimizing an auxiliary function
\begin{equation}
\label{PSI}
\Psi(\{a_{i}\},\{\lambda_{i}\}) =\chi^{2}(\{a_{i}\})+\sum_{j} \lambda_{j}\,\textit{O}_{j}(\{a_i\})
\end{equation}
with respect to the set of fit parameters $\{a_i\}$ for fixed values of the Lagrange 
multipliers $\{\lambda_{j}\}$.
Each multiplier is related to a specific observable 
$\textit{O}_{j}$, and the choice $\lambda_{j}=0$ corresponds to the best fit. 
By repeating the minimization procedure multiple times with respect to $\Psi(\{a_{i}\},\{\lambda_{i}\})$
for different, fixed values of $\{\lambda_{j}\}$ one can map out precisely 
how the fit to data deteriorates when the expectation for the 
observable $\textit{O}_{j}$ is forced to change from its optimum value.
The procedure also determines the range of variation of the observable 
within a given choice of tolerance criterion. 

Finally, in order to make comparisons with the Lagrange multiplier method in 
the following, and also to supplement the Monte Carlo data sampling 
with low-$x$ pseudo-data as described above, we adopt the procedure 
described in Ref.~\cite{Martin:2009iq} for defining our $68\%$ C.L.\ interval.
More specifically, we choose the maximal variation of any quantity of interest 
that keeps the increase of the partial contribution to the effective $\chi^{2}$ function 
of every experiment included in the fit, $\chi^2_n$, at most proportional to the increase 
expected for a $\chi^2$-distribution with $N$ degrees of freedom from
the most probable value $\xi_{50}$ to the $68^{th}$ percentile $\xi_{68}$, i.e.,
we demand that
\begin{equation}
\chi^2_n \le \left( \frac{\chi^2_{n,0}}{\xi_{50}}\right)\,\xi_{68}\,.
\end{equation}
Here, $\chi^2_{n,0}$ is the best fit value for $\chi^2_{n}$, and $\xi_{68}$ is defined by
\begin{equation}
\int_0^{\xi_{68}} d\chi^2\, P_N(\chi^2)=0.68\,,
\end{equation}
with
\begin{equation}
P_N(\chi^2)=\frac{(\chi^2)^{\frac{N}{2}-1}\, e^{-\chi^2/2}}{2^{\frac{N}{2}}\, \Gamma\left(\frac{N}{2}\right)}\,,
\end{equation}
and where $N$ denotes the number of data points in the $n$-th data set under consideration.

\section{Results}
%
In this section we present and discuss the results of our Monte Carlo sampling strategy for helicity 
PDFs and their uncertainties and compare them to those obtained in the 
DSSV14 analysis based on traditional fitting. 

\begin{figure}[th!]
\vspace*{0.2cm}
\begin{center}
\hspace*{-0.65cm}
\epsfig{figure=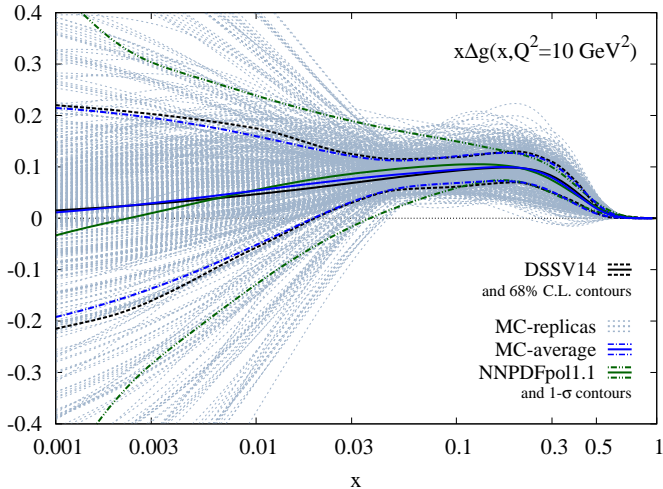,width=0.54\textwidth}
\end{center}
\vspace*{-.5cm}
\caption{The ensemble of replicas (dotted blue lines) for the NLO gluon helicity density $\Delta g(x,Q^2)$ 
at $Q^{2}=10\,$GeV$^2$ shown along with its statistical average (solid blue line)
and variance (dot-dashed blue lines). 
The corresponding results from the DSSV14 fit (black lines) and the NNPDFpol1.1 analysis 
(green lines) are shown for comparison; see text.
\label{fig:gluon}}
\end{figure}

We start with the phenomenologically most interesting quantity, the helicity gluon density $\Delta g(x,Q^2)$ at NLO
accuracy, which is shown in Fig.~\ref{fig:gluon} as a function of $x$ at a representative scale of
$Q^2=10\,\mathrm{GeV}^2$. Given is our newly obtained ensemble of replicas along with 
its statistical average (solid blue line) and variance (dot-dashed blue lines), 
representing the best fit $\Delta g$ and the 1-$\sigma$ uncertainty interval, respectively. 
The result of the original DSSV14 best fit and the contour covering the corresponding
$68\%$ C.L.\ interval, computed with the Lagrange multiplier technique and
tolerance criterion outlined at the end of Sec.~\ref{sec:setup}, 
are illustrated for comparison by the solid and dashed black lines, respectively.
%
%
%

\begin{figure*}[th!]
\vspace*{0.2cm}
\begin{center}
\hspace*{-0.3cm}
\epsfig{figure=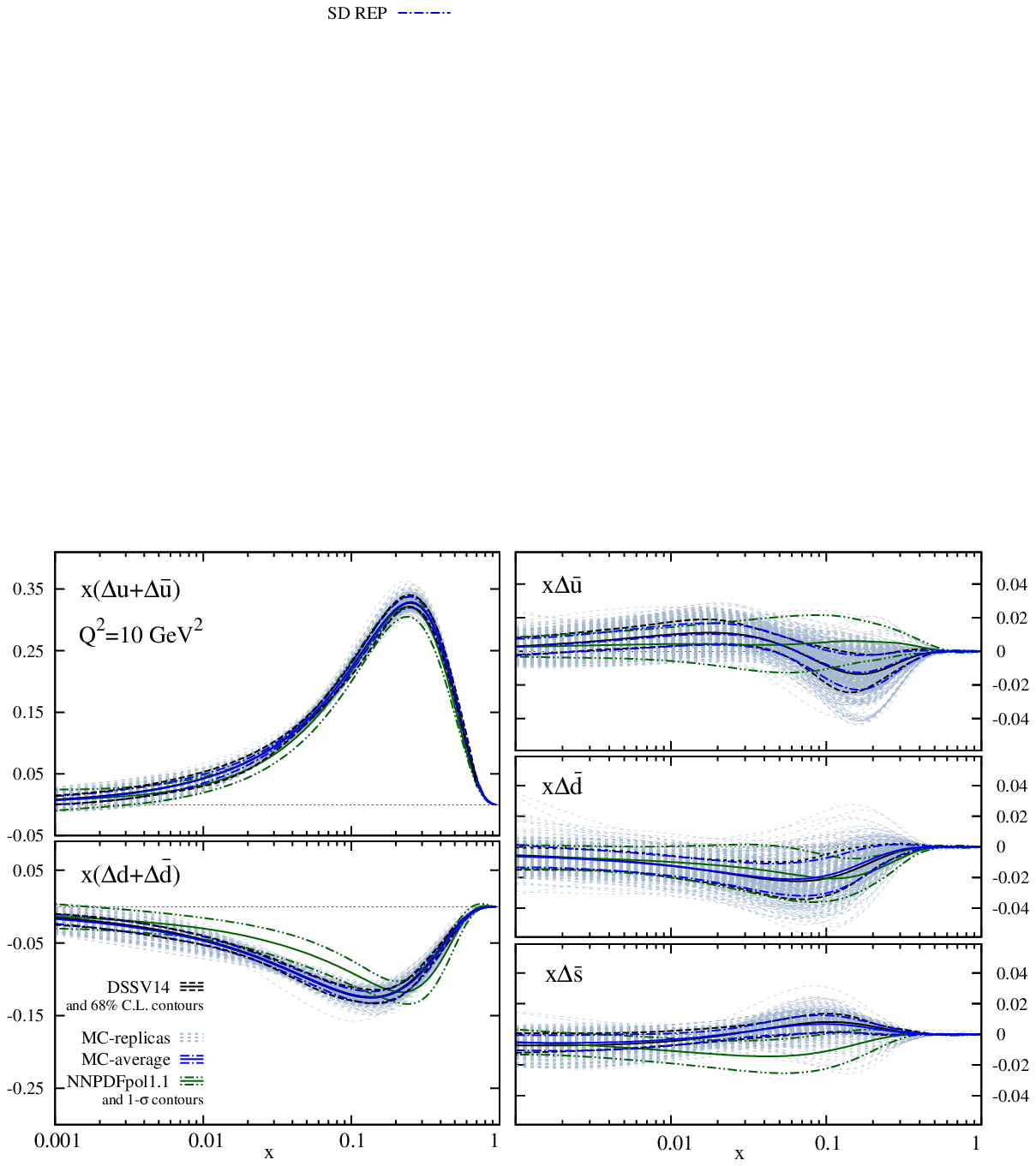,width=1.02\textwidth}
\end{center}
\vspace*{-.5cm}
\caption{Same as Fig.~\ref{fig:gluon} but now showing our results for the quark and antiquark helicity PDFs 
at $Q^{2}=10\,\mathrm{GeV}^2$ in comparison to the analyses of DSSV14 and NNPDFpol1.1. 
\label{fig:quarks}}
\end{figure*} 
As can be noticed, most of the replicas resemble closely the DSSV14 best fit down to about $x \simeq 0.05$
where a large number of them starts to diverge ever more rapidly for decreasing momentum fractions,
resulting in a significant broadening of the uncertainty band.
This noticeable change in the behavior of the replicas is closely related to the range 
of $x$ predominantly probed by RHIC $pp$ data, which deliver the most stringent 
and direct constraints on the gluon polarization to date.
The statistical average of the ensemble of our 1000 replicas closely matches 
the DSSV14 best fit, but, as has to be expected, the agreement is not 
perfect due to the increased flexibility in the functional form (\ref{eq:glue-input})
adopted in the present analysis.
It is interesting to notice that also the 1-$\sigma$ variance of the replicas approximates 
rather closely the $68\%$ C.L.\ uncertainties coming from the Lagrange multiplier method.
This is a nontrivial, and perhaps even unexpected result in view of 
the large tolerances $\Delta \chi^2$ of the order of $10$ to $15$ units 
that are allowed for in the uncertainty estimates for DSSV14 based on Lagrange multipliers. 
Of course, the Monte Carlo replicas and, hence, their corresponding 1-$\sigma$ variance, are
designed to effectively cope with neglected uncertainties, like those related to theoretical 
approximations and assumptions, that are not accounted in the effective $\chi^2$ function
and which also cause the large tolerances adopted in the Lagrange multiplier method.
 
As a further comparison, Fig.~\ref{fig:gluon} also incorporates the results (green lines)
from the NNPDFpol1.1 analysis \cite{Nocera:2014gqa}
which is based on a Monte Carlo sampling of spin-dependent DIS data 
and a largely unbiased interpolation of the $x$-dependence of helicity PDFs by a neural network. 
It also includes information on inclusive jet and $W$-boson production from RHIC, but neither SIDIS data
nor spin asymmetries for inclusive neutral pion production at RHIC are used so far, 
both of which play an important role in the DSSV14 global analysis.
Nevertheless, the results are very much compatible and show a remarkable agreement for both the
central values and uncertainty estimates in the $x$-range constrained by jet and DIS data.
At lower values of $x$, the uncertainties in $\Delta g(x,Q^2)$ are largest for the NNPDFpol1.1 analysis.
This observation can be explained at least in part by the missing information from 
neutral pion production at RHIC, which constrains $\Delta g$ down to somewhat lower values of $x$ than
jet data alone \cite{deFlorian:2014yva}.

\begin{figure}[b!]
\vspace*{-0.5cm}
\begin{center}
\hspace*{-0.35cm}
\epsfig{figure=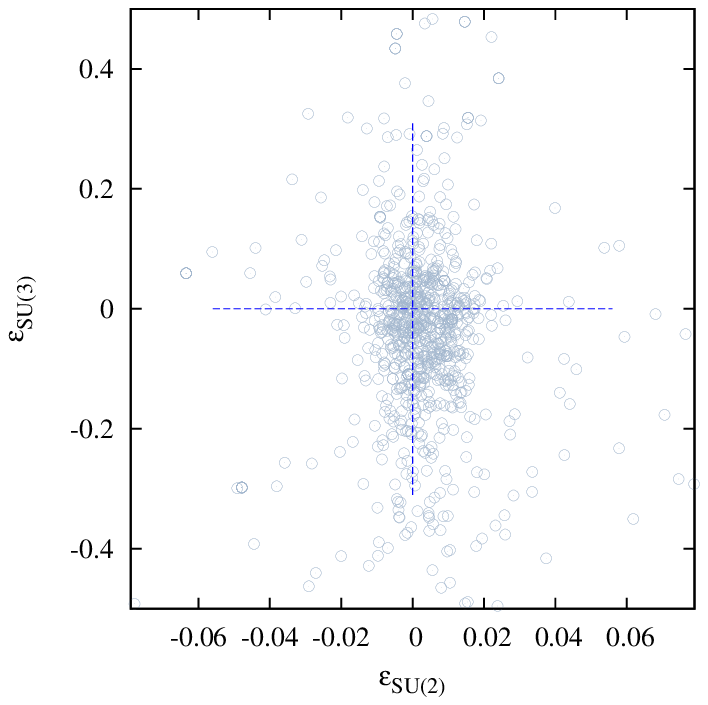,width=0.53\textwidth}
\end{center}
\vspace*{-.5cm}
\caption{The ${\mathrm{SU}}(2)$ and ${\mathrm{SU}}(3)$ symmetry breaking parameters in Eqs.~(\ref{eq:su2}) and
(\ref{eq:su3}) for our ensemble of helicity PDF replicas.
\label{fig:su}}
\end{figure} 
Similar observations can be made about the quark and antiquark helicity distributions, which
can be found in Fig.~\ref{fig:quarks}. As in Fig.~\ref{fig:gluon}, we show the newly obtained ensembles of replicas
for $\Delta u+\Delta \bar{u}$ and $\Delta d+\Delta \bar{d}$ (left-hand panels) and
$\Delta \bar{u}$, $\Delta \bar{d}$, $\Delta \bar{s}\equiv\Delta s$ (right-hand panels),
their statistical averages and variances. Again, for comparison, results stemming from
the analyses by DSSV14 and NNPDFpol1.1 are presented as well.

As can be inferred from the left-hand panels, the flavor combination $\Delta u+\Delta \bar{u}$ in particular,
but also $\Delta d+\Delta \bar{d}$, both of which are probed by DIS data, 
exhibit the smallest uncertainties of all helicity densities. 
Most of the replicas are closely concentrated around their average in the medium-to-large $x$ region where
the valence quark contributions to $\Delta q+\Delta \bar{q}$, $q=u,\,d$, are dominant.
Here, the relative errors amount to about $5\%$ and $20\%$ for $\Delta u+\Delta \bar{u}$ and $\Delta d+\Delta \bar{d}$,
respectively. The dispersions of replicas become more pronounced for smaller momentum fractions, where sea quarks rule,
with relative uncertainties increasing to about $100\%$, which is still significantly smaller than for $\Delta g$ shown
in Fig.~\ref{fig:gluon}.
In general, the constraints on the sea quark distributions are rather weak in the entire range of $x$ probed by the data
as can be gathered from the right-hand panels of Fig.~\ref{fig:quarks}. They receive their constraints mainly from
SIDIS data that are less precise than fully inclusive measurements and suffer from additional theoretical ambiguities 
from fragmentation functions.  

As for $\Delta g$, the agreement with the results from the traditional global analysis of the DSSV group
is very good for all quark flavors both for the average, i.e., best fit, and the uncertainty bands. Again,
the latter are obtained with the Lagrange multiplier method assuming inflated tolerance criteria for $\Delta \chi^2$.
The results from NNPDFpol1.1 compare less favorably to our results except for  $\Delta u+\Delta \bar{u}$ and, perhaps,
$\Delta \bar{d}$. However, here it should be kept in mind that the NNPDF group so far does not include any
SIDIS data in their analysis. On the other hand, they achieve some flavor discrimination through reweighting
their replicas with recent results on $W^{\pm}$-boson single-spin asymmetries
from RHIC \cite{Aggarwal:2010vc,Adare:2010xa,Adamczyk:2014xyw},
which are included neither in DSSV14 nor in the present analysis. 
This likely explains the differences observed for $\Delta d+\Delta \bar{d}$ and $\Delta \bar{u}$.
Our results for $\Delta s$ are largely driven by SIDIS data with observed charged kaons in the final-state
\cite{deFlorian:2008mr,deFlorian:2009vb,deFlorian:2014yva} while
for NNPDFpol1.1 the only constraint is derived from the baryonic semi-leptonic $\beta$-decay parameters,
to which we turn next, which prefer a negative $\Delta s$. 

The often adopted constraints on the first moments of the total quark helicity densities  
from baryonic semi-leptonic $\beta$-decay parameters $F$ and $D$, i.e., SU(2) and SU(3) symmetry arguments,
deserve some further scrutiny and discussion. Clearly, violations of SU(3) symmetry are expected at some level; 
see, e.g., Refs.~\cite{Lichtenstadt:1995xk,Savage:1996zd,Bass:2009ed} and references therein.
Rather than imposing the symmetry constraints at face value, deviations were allowed and measured 
in terms of two additional fit parameters $\varepsilon_{{\mathrm{SU}}(2)}$ and $\varepsilon_{{\mathrm{SU}}(3)}$ 
in all previous DSSV analyses \cite{deFlorian:2008mr,deFlorian:2009vb,deFlorian:2014yva}.
More specifically, the $F$ and $D$ values were related to the 
first moments by
\begin{eqnarray}
\label{eq:su2}
\Delta \Sigma_u - \Delta \Sigma_d &=&
(F+D)\, [1+\varepsilon_{{\mathrm{SU}}(2)}], \\ \nonumber \\
\label{eq:su3}
\Delta \Sigma_u + \Delta \Sigma_d-2 \Delta \Sigma_s &=&
(3F-D)\, [1+\varepsilon_{{\mathrm{SU}}(3)}],
\end{eqnarray}
where
\begin{eqnarray}
\label{eq:firstmom1}
\Delta \Sigma_f 
 \equiv  \int_0^1 \left[ \Delta f_i + \Delta \bar{f}_i \right] (x,\mu_0)\,dx 
\,,
\end{eqnarray}
with $F+D=1.269\pm0.003$ and $3F-D=0.586\pm0.031$ (see, e.g., Ref.~\cite{ref:hermes-a1pd})
at the input scale $\mu_0=1\,\mathrm{GeV}$ of the DSSV analysis. Note that
both relations (\ref{eq:su2}) and (\ref{eq:su3}) are renormalization group invariants,
i.e., are scale independent. In practice, the free fit parameters
$\varepsilon_{{\mathrm{SU}}(2)}$ and $\varepsilon_{{\mathrm{SU}}(3)}$ 
substitute the normalizations $N_{u+\bar{u}}$ and $N_{d+\bar{d}}$ of the corresponding 
quark distributions in Eq.~\ref{eq:pdf-input}, which otherwise could have been fixed 
by $F$ and $D$.

Also in our present analysis, the two combinations (\ref{eq:su2}) and (\ref{eq:su3})
including the $F$ and $D$ constants are taken as two additional data points, i.e., are 
included in the effective $\chi^2$ function and shifted around their central 
values as any other measurement when determining the ensemble of data and PDF replicas. 
Consequently, each PDF replica inherits its own values for $\varepsilon_{{\mathrm{SU}}(2)}$ and 
$\varepsilon_{{\mathrm{SU}}(3)}$ that quantify the departure from 
${\mathrm{SU}}(2)$ and ${\mathrm{SU}}(3)$ symmetry, respectively.

Figure~\ref{fig:su} illustrates the distribution of the two symmetry breaking parameters 
for our ensemble of replicas. We obtain $\varepsilon_{{\mathrm{SU}}(2)}=0.000 \pm 0.056$ and 
$\varepsilon_{{\mathrm{SU}}(3)}=0.000 \pm 0.311$. The average values are compatible with zero,
which mostly reflects the fact that large departures from SU(3) symmetry come with a penalty
in $\chi^2$ in our approach. Interestingly, the variances are somewhat larger
than expected from the experimental uncertainties of the $F+D$ and $3F-D$ values alone,
which shows the influence of the DIS and, especially, the SIDIS data. 
In this way, our ensemble of helicity PDFs replicas and, most importantly,
any uncertainties for observables obtained with them, explore a fairly wide range of
symmetry breaking possibilities. We note that in Ref.~\cite{Ethier:2017zbq} a simultaneous 
determination of helicity parton densities and fragmentation function was performed, in which
the values for the triplet and octet axial charges were freely fitted. Our replicas
necessarily have a larger octet charge than that found in~\cite{Ethier:2017zbq}, although their
spread is not too different from the uncertainty quoted there.

\section{reweighting Applied to Dijet Data}
%
As a first application to our newly generated set of helicity PDF replicas,
we apply the reweighting technique \cite{Ball:2010gb,Ball:2011gg} 
to estimate the impact of recent STAR data for dijets measured in polarized $pp$ collisions at a
center-of-mass system energy of $\sqrt{S}=200\,\mathrm{GeV}$ and for central and forward
pseudo-rapidity configurations of the two jets \cite{Adamczyk:2016okk,Adam:2018pns}.
The corresponding double-spin asymmetry $A_{\mathrm{LL}}$ will be evaluated at NLO accuracy adopting the
calculation in Ref.~\cite{deFlorian:1998qp}.
At variance with the largely analytical results available for single-inclusive
observables such as the high-transverse momentum 
production of pions \cite{Jager:2002xm} and jets \cite{Jager:2004jh},
that are already routinely used in fits, calculations for dijet production incorporate
time-consuming, numerical phase-space integrations making their practical
implementation in a global analysis more cumbersome.
As in all DSSV-type global analyses, exact NLO expressions are implemented 
most efficiently in Mellin moment space, see Ref.~\cite{deFlorian:2009vb}
for an outline of the method. This also works very well for dijet production.

We note that the full set of STAR dijet data \cite{Adamczyk:2016okk,Adam:2018pns} 
has not been used in any global analysis of helicity PDFs so far,
although the central data set \cite{Adamczyk:2016okk} was included by reweighting
in Ref.~\cite{Nocera:2017wep}. Exploring the relevance of both data sets in
constraining helicity PDFs further is, we believe, a timely and important exercise.
Dijet data receive their potential relevance for PDF determinations
from probing parton momentum fractions in a more controlled way than single-inclusive probes,
for which the information on $x$ is integrated over a large range.
This is achieved by selecting distinctly different dijet topologies, 
defined by the pseudo-rapidities of the two observed jets.
It is expected that dijet data will complement especially the information available on the
gluon helicity density, coming so far mainly from single-inclusive jet and neutral pion production measurements
at RHIC. This is particularly relevant in order to check to which extent the 
new dijet data corroborate and perhaps ameliorate the evidence for a sizable positive gluon polarization 
at medium-to-large values of $x$ based on single-inclusive measurements and 
reported in Refs.~\cite{deFlorian:2014yva,Nocera:2014gqa}.

The reweighting technique \cite{Ball:2010gb,Ball:2011gg} allows one to incorporate 
consistently the information provided
by a new set of data into an existing ensemble of PDF replicas without the need of 
refitting them, but preserving the statistical rigor of its extraction. 
The usefulness of the method in the context of PDF determinations has already been 
successfully demonstrated in different applications, see, for instance, 
Refs.~\cite{Nocera:2014gqa,Armesto:2013kqa,Paukkunen:2014zia,Borsa:2017vwy}.

Using Bayesian inference, it is possible to update the original probability distribution 
of an ensemble of PDF replicas to one that accounts for the information contained 
in a new measurement~\cite{Ball:2010gb}. To this end, one assigns a new weight $w_k < 1$ to each replica 
which measures its agreement with the new data. 
The Bayesian reweighting is fully equivalent to a refit including the additional set of data
as long as the impact of the new experimental information is not too significant, 
for instance, by constraining some aspect of the PDFs that was largely undetermined before. 
Such a scenario would lead to a very large number of replicas with essentially vanishing weights $w_k$,
making a full refit inevitable.
Next, we present the effect of reweighting our ensemble of helicity PDFs replicas 
with data sets from the STAR collaboration \cite{Adamczyk:2016okk,Adam:2018pns}
corresponding to different dijet rapidity configurations, using them both
one-by-one and combined into a single data set.

\begin{figure}[t!]
\vspace*{0.2cm}
\begin{center}
\hspace*{-0.65cm}
\epsfig{figure=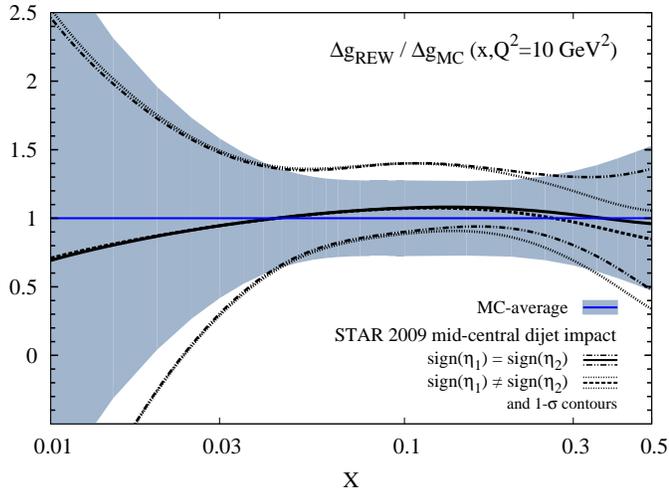,width=0.54\textwidth}
\end{center}
\vspace*{-.5cm}
\caption{Impact of STAR mid-central rapidity dijet data \cite{Adamczyk:2016okk} 
on the gluon helicity distribution $\Delta g(x,Q^2)$ 
as a function of $x$ at $Q^2=10\,\mathrm{GeV}^2$. 
The plot shows the averages and variances of our replicas for $\Delta g$ 
after being reweighted with same and opposite sign configurations of the two jets (different line styles),
see text, normalized to the average before reweighting. The shaded band gives the
1-$\sigma$ uncertainties for the ensemble of $\Delta g$ replicas before reweighting,
normalized to the average.
\label{fig:gluon2}}
\end{figure} 
We start by showing in Fig.~\ref{fig:gluon2} the impact on $\Delta g(x,Q^2)$ from
reweighting our Monte Carlo replicas with the STAR 2009 results \cite{Adamczyk:2016okk} for mid-central 
(''barrel'' detector) pseudo-rapidity configurations of the two jets, $-0.8\le \eta \le 0.8$, 
in the relevant range of $x$ predominantly probed by the data.
The experimental results are provided separately for two topologies 
where both observed jets are either reconstructed in the same or in opposite hemispheres, 
henceforth labeled as  ''same-sign'' and  ''opposite-sign'' configurations, respectively.
We perform an independent reweighting for each of these two subsets of data
and show the resulting averages and variances normalized to the statistical average of 
$\Delta g$ replicas before reweighting. 
To facilitate the comparison to the results shown in Fig.~\ref{fig:gluon},
the reweighted $\Delta g$ is presented at a common scale of $Q^2=10\,\mathrm{GeV}$,
but the NLO calculations for dijets \cite{deFlorian:1998qp} are performed at the scale of the 
respective dijet invariant mass for each point, including all relevant kinematical cuts 
made in experiment \cite{Adamczyk:2016okk}. 
The uncertainty estimates do not include contributions coming from the factorization scale dependence or 
those associated with the choice of unpolarized PDFs used to normalize the asymmetries.
As a reference, the shaded band in Fig.~\ref{fig:gluon2} gives the 1-$\sigma$ uncertainty for the ensemble of 
$\Delta g$ replicas before reweighting, normalized to the average.

As can inferred from comparing the uncertainty bands in Fig.~\ref{fig:gluon2} before and after reweighting,
the most significant effect is found around $x\simeq 0.15$ with a noticeable reduction of the width of the band. 
In addition, an approximately $10\%$ increase in the average $\Delta g(x,Q^2)$ as compared to the original 
distribution is found here after reweighting.
Both sets of data show a very similar trend for $x\lesssim0.15$, but the opposite-sign configuration 
prefers somewhat less polarization towards larger values of $x$. 
At the lower end of the $x$-range shown in Fig.~\ref{fig:gluon2}, i.e., for $x \lesssim 0.05$, 
the reweighted averages start to drop below the original $\Delta g(x,Q^2)$, but at the same time
the uncertainty bands remain essentially unchanged. This suggests, as one can anticipate already
from kinematical considerations, that the dijet data sets do not lead to any further constraints in this region. 
The behavior at small $x$ is most likely induced by the original data in the fit to compensate for the slight
increase around $x\simeq 0.15$ in order to keep the first moment roughly constant.
Here, a complete re-analysis without resorting to the reweighting method might
shed more light on this observation.

It should be noted that the number of replicas with a non-negligible weight after
reweighting, see Ref.~\cite{Ball:2010gb} for details on this criterion, 
is large for both sets of data, amounting to 783 and 749 members for same-sign and opposite-sign configurations, respectively.
This also underpins the general observation that the information on helicity
PDFs, in particular, on $\Delta g$, contained in the dijet data is fully consistent
with what has been obtained already in the DSSV14 global analysis based on single-inclusive RHIC data.

A second set of STAR dijet data \cite{Adam:2018pns} contains configurations with at least one
jet reconstructed at forward pseudo-rapidities (''endcap'' detector), covering the
range $0.8\le \eta\le 1.8$. Together with the possibility
that one of the jets is detected in the barrel detector, i.e., at mid rapidity $|\eta|\le 0.8$, this gives
us three additional data sets for a reweighting exercise, which we label
as ''west barrel-endcap'', ''east-barrel-endcap'', and ''endcap-endcap'',
where ''west'' denotes the hemisphere with $\eta>0$. 
%
\begin{figure}[t!]
\vspace*{0.2cm}
\begin{center}
\hspace*{-0.65cm}
\epsfig{figure=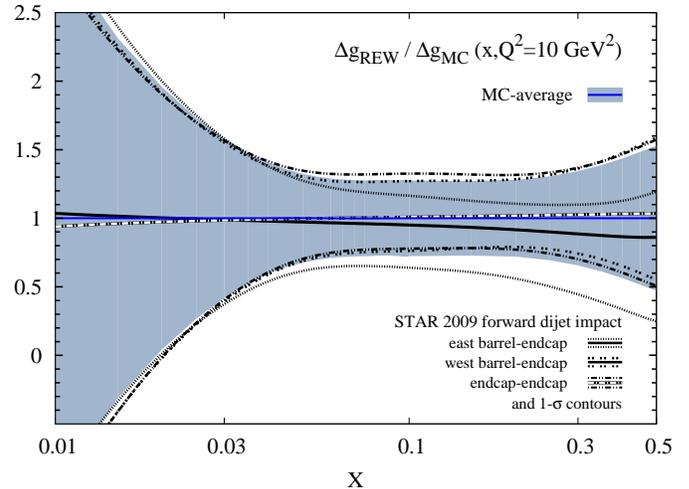,width=0.54\textwidth}
\end{center}
\vspace*{-.8cm}
\caption{Same as in Fig.~\ref{fig:gluon2} but now for the three subsets of
dijet configurations from the STAR collaboration \cite{Adam:2018pns}
with at least one jet at forward (endcap) rapidities, see text.
\label{fig:gluon3}}
\end{figure} 

The results of the reweighting can be found in Fig.~\ref{fig:gluon3}.
Compared to the data with two jets at mid rapidity shown in Fig.~\ref{fig:gluon2},
the impact of the forward jets in the reweighting procedure 
is considerably less pronounced. While the two dijet topologies 
west barrel-endcap and endcap-endcap produce almost no effect on the
reweighted averages and variances, the data on the east-barrel-endcap 
configuration show some trend towards a smaller average gluon polarization for $x\gtrsim 0.1$
but at the same time with almost no changes in the width of the corresponding uncertainty band,
making it inconclusive.
In general, the much weaker impact of the forward dijet configurations can be  
associated with the comparatively larger experimental uncertainties of these sets of data.
This is also reflected in the large number of replicas with a non-negligible weight after
reweighting: 857, 964, and 956 for the east-barrel-endcap, west barrel-endcap 
and endcap-endcap configurations, respectively, close to the 1000 original replicas we started from.
Upon closer inspection, the central values of the measured double-spin asymmetries 
for the west barrel-endcap and the endcap-endcap dijet configurations
suggest a trend for a larger gluon polarization at $x \sim 0.1$, but the sizable experimental errors 
undermine their impact in the reweighting process. 
We have explicitly verified that scaling the experimental 
errors artificially down in the computation of the new weights $w_k$ results in an
increase of the gluon polarization.

%
\begin{figure}[t!]
\vspace*{0.2cm}
\begin{center}
\hspace*{-0.65cm}
\epsfig{figure=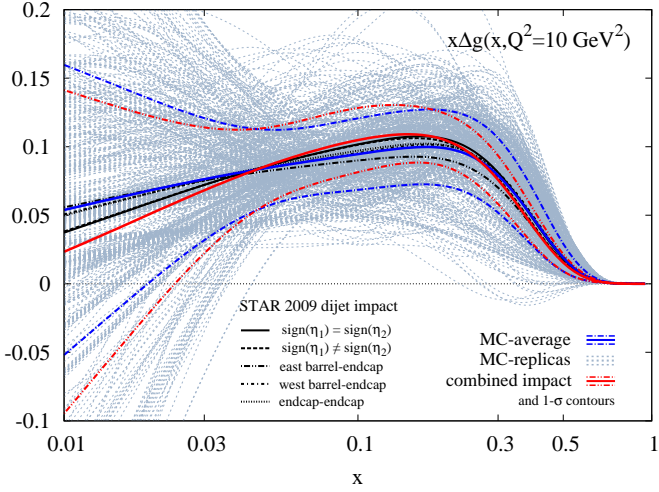,width=0.54\textwidth}
\end{center}
\vspace*{-.5cm}
\caption{Combined effect of the STAR 2009 mid-central and forward dijet sets of 
data \cite{Adamczyk:2016okk,Adam:2018pns} on the gluon helicity distribution
in the relevant range of $x$ for $Q^2=10\,\mathrm{GeV}^2$.
Shown are the average and variance (1-$\sigma$ contour) 
before (blue lines) and after (red lines) reweighting, the original ensemble of
replicas, and the individual averages from each data set (black lines) as given 
in Figs.~\ref{fig:gluon2} and \ref{fig:gluon3}.
\label{fig:gluon4}}
\end{figure} 
The combined impact of all STAR 2009 dijet data sets
on the gluon helicity distribution $\Delta g(x,Q^2)$
can be found in Fig.~\ref{fig:gluon4} in the relevant range of $x$ for $Q^2=10\,\mathrm{GeV}^2$.
Shown are the average and variance before (blue lines) and after (red lines) 
reweighting. For reference, we also give the five average
gluon helicity densities (black lines) from the individual reweighting exercises
discussed in Figs.~\ref{fig:gluon2} and \ref{fig:gluon3}.
As can be seen, the overall impact of the combined set of dijet data 
is a very moderate increase of the gluon polarization in the range $0.05\lesssim x \lesssim 0.2$,
well within the uncertainty of the DSSV14 replicas, and a sizable reduction of the width 
of the 1-$\sigma$ uncertainty band, most noticeable for $x \gtrsim 0.2$. 
This nicely confirms both the evidence for a positive gluon polarization at intermediate 
values of $x$ first demonstrated in Ref.~\cite{deFlorian:2014yva}
and the anticipated impact of the dijet probe on $\Delta g(x,Q^2)$.

For future reference and to illustrate again the impact and consistency of the dijet data,
we quote here some representative values and 1-$\sigma$ uncertainties for truncated
moments of the gluon helicity density, $\int_{x_{\min}}^1 \Delta g(x,Q^2)\, dx$, at $Q^2=10\,\mathrm{GeV}^2$. 
For $x_{\min}=0.01$ we obtain $0.309\pm 0.109$ and $0.296\pm 0.108$ before and after reweighting, respectively.
Likewise, for $x_{\min}=0.1$ the corresponding numbers read $0.133\pm 0.035$ and $0.126\pm 0.023$.
It should be noted that the values before reweighting are fully consistent with those
obtained from the Lagrange multiplier method in the original DSSV14 analysis.

\begin{figure}[th!]
\vspace*{-0.7cm}
\begin{center}
\hspace*{-0.49cm}
\epsfig{figure=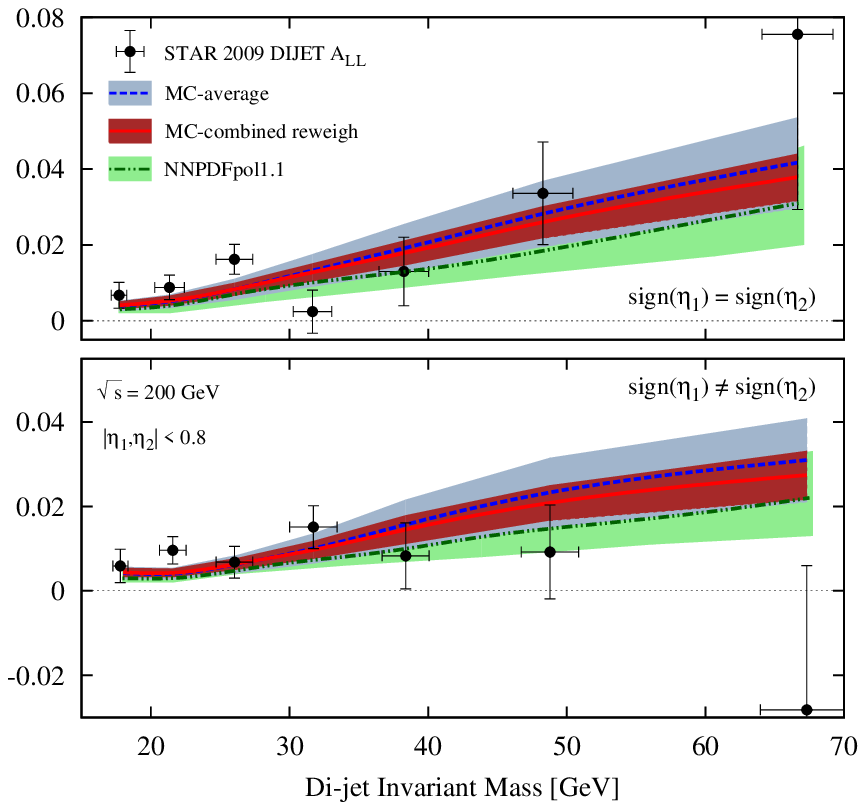,width=0.71\textwidth}
\end{center}
\vspace*{-.5cm}
\caption{The STAR 2009 mid-central dijet double spin asymmetries $A_{\mathrm{LL}}$ \cite{Adamczyk:2016okk}
as a function of the invariant mass of the jet pair compared to the averages of the original and MC-reweighted
ensemble of our replicas. The shaded bands illustrate the respective variances.
Also shown are the results obtained with the NNPDFpol1.1 set of helicity PDFs.
\label{fig:gluon5}}
\vspace*{-0.4cm}
\begin{center}
\hspace*{-0.49cm}
\epsfig{figure=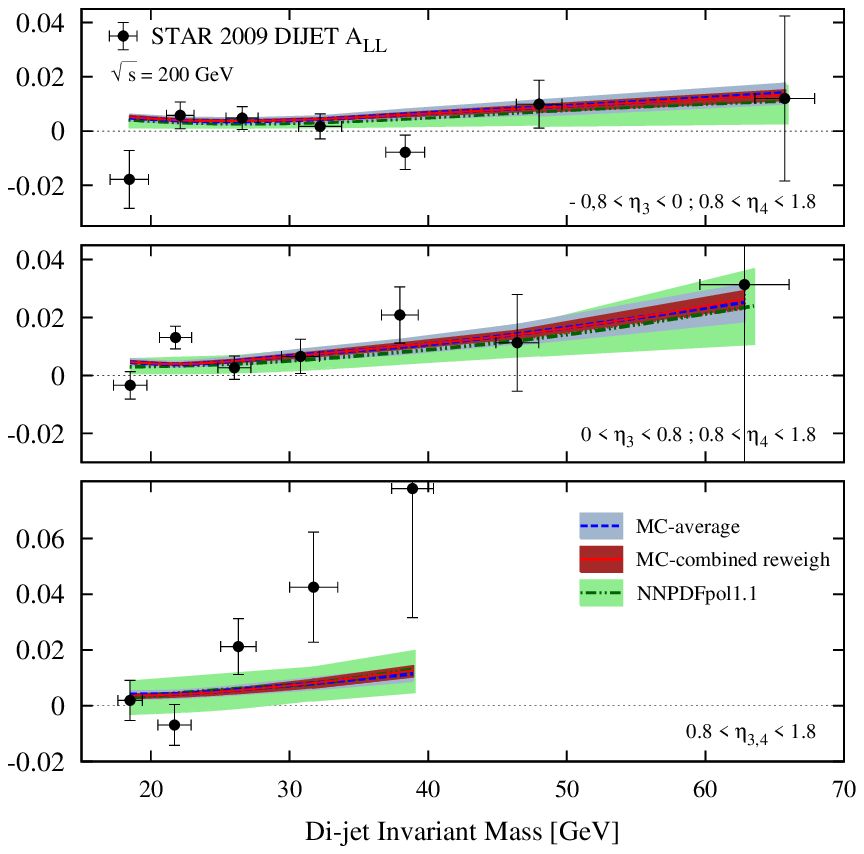,width=0.71\textwidth}
\end{center}
\vspace*{-.5cm}
\caption{As in Fig.~\ref{fig:gluon5} but now for the three sets of STAR 2009 forward dijet spin asymmetries \cite{Adam:2018pns}.
\label{fig:gluon6}}
\end{figure} 
Finally, in Figs.~\ref{fig:gluon5} and \ref{fig:gluon6} we compare the actually measured
double-spin asymmetries $A_{\mathrm{LL}}$ from the STAR collaboration 
for the mid-central and forward dijet configurations \cite{Adamczyk:2016okk,Adam:2018pns}, 
respectively, as a function of the invariant mass
of the jet pair with the averages of the original and reweighted
ensemble of our replicas. The shaded bands illustrate the corresponding 1-$\sigma$ uncertainty bands.
As is expected, the changes in the central values for $A_{\mathrm{LL}}$ before and after reweighting
are rather moderate and both results are compatible with the data. However, there is a quite noticeable reduction
in the width of the uncertainty bands for all sets of dijet data after the reweighting procedure.
Note that the data points at lower invariant mass have the smallest uncertainties and
hence the biggest impact in the reweighting procedure. 
For comparison,  Figs.~\ref{fig:gluon5} and \ref{fig:gluon6} also contain the result
of a NLO calculation of  $A_{\mathrm{LL}}$ utilizing the set of replicas from NNPDFpol1.1.  
In general, the use of NNPDFpol1.1 yields smaller double-spin asymmetries but still consistent with
the STAR data within their larger uncertainty estimates. Interestingly, the description of the 
data set with both jets forward remains overall poor within the present theoretical calculations.
It may be especially interesting here to explore the uncertainties related to the choice
of factorization or renormalization scales, as well as the influence of the spin-averaged PDFs
in the denominator of the spin asymmetry.

\section{Conclusions}
%
In this paper we have explored the feasibility of combining a Monte Carlo sampling strategy
with the traditional fitting approach adopted by the DSSV group to extract helicity parton densities 
from a global QCD analysis at NLO accuracy. To facilitate the comparison between the two methods,
the data sets analyzed, as well as other fit inputs were chosen to be identical to 
those of the DSSV14 analysis.

The main advantages of the Monte Carlo approach are, on the one hand, the availability of a large set of PDF
replicas that allows one to straightforwardly estimate and propagate the PDF uncertainties to other observables
solely by statistical means, i.e., by computing the average and variance, without the need of 
an effective $\chi^2$ function to assess the agreement with data.
On the other hand, the standard interpolation for the dependence of PDFs on 
the momentum fraction $x$ with fixed but flexible functional forms for each parton flavor
allows for the use of numerically efficient calculational tools, for instance, based on Mellin moment space, 
to compute NLO QCD cross sections without the need of approximations.
Having explicit parametrizations for each of the replicas at hand might also be convenient in understanding the
observed features imprinted on them by the data.
In addition, the availability of replicas of PDFs opens up the possibility of quickly implementing new
sets of data with the reweighting technique to study the impact on the PDFs without the need of refitting them.

The results obtained with our combined approach based on Monte Carlo replicas agree fairly well with those 
from the standard DSSV14 global analysis both for the optimum fit and the uncertainty estimates.
While calculating the variance of the replicas avoids the adoption of a tolerance criterion, 
the DSSV14 approach is based on some inflated $\Delta \chi^2$ to account for sources
of uncertainties that are not necessarily included in the effective $\chi^2$ function but which
become apparent when judging the agreement of the fit to the various sets of data adopted in the analysis.

As a first application of our new set of helicity PDF replicas, we have invoked the 
reweighting procedure to reveal the impact of the recent STAR dijet data for
different jet topologies on the determination of the momentum fraction dependence of the
gluon helicity distribution. 
We find that, with the exception of data with two forward jets, which have comparatively
large uncertainties, the double-spin asymmetries for all dijet topologies
are in very good agreement with previous RHIC measurements and
corroborate and strengthen the evidence for a sizable positive gluon polarization 
at medium-to-large values of $x$ discussed in the literature.

\section*{Acknowledgments}
%
We warmly acknowledge Emanuele R.\ Nocera for enlightening discussions on the Monte 
Carlo sampling approach, Elke C.\ Aschenauer and Carl Gagliardi for their help with STAR 
dijet data. 
This work was supported in part by CONICET and ANPCyT.


\end{document}